\documentclass[sigconf]{acmart}
\copyrightyear{2022}
\acmYear{2022}
\setcopyright{acmcopyright}\acmConference[ICSE '22]{44th International Conference on Software Engineering}{May 21--29, 2022}{Pittsburgh, PA, USA}
\acmBooktitle{44th International Conference on Software Engineering (ICSE '22), May 21--29, 2022, Pittsburgh, PA, USA}
\acmPrice{15.00}
\acmDOI{10.1145/3510003.3510141}
\acmISBN{978-1-4503-9221-1/22/05}

\usepackage[english]{babel}

\usepackage[font=small, textfont=bf]{caption}
\usepackage{amsmath}
\usepackage{graphicx}
\usepackage{booktabs}
\usepackage{listings}
\usepackage{algorithm}
\usepackage{subcaption}
\usepackage[noend]{algpseudocode}
\usepackage{tikz}
\usepackage{longfbox}

\definecolor{mred}{rgb}{.80,.12,.30}
\definecolor{grey}{rgb}{0.5,0.5,0.5}
\definecolor{purple2}{rgb}{.75,0,.85}
\definecolor{pistachio}{rgb}{0.38, 0.57, 0.25}
\definecolor{steelblue}{rgb}{.20,.35,.90}

\newcommand{\ie}[1]{(i.e., #1)}
\newcommand{\eg}[1]{(e.g., #1)}

\newcommand{\totalbugs}[0]{57}
\newcommand{\DISTINCTBUGS}[0]{30}
\newcommand{\toolName}[0]{\texttt{TOGA}}

\newcommand{\repo}[0]{\url{https://github.com/microsoft/toga}}

\input{sections/code_highlight}
\newcommand{\TODO}[1]{\textcolor{red}{{\textbf{TODO: #1}}}}

\newcommand{\parheader}[1] {\noindent\textbf{#1}}

\title{\toolName{}: A Neural Method for Test Oracle Generation}

\begin{document}

\author{Elizabeth Dinella}
\authornote{Performed this work while interning at Microsoft. Equal contributor.}          
\affiliation{
  \institution{University of Pennsylvania}            
}
\email{edinella@seas.upenn.edu}          

\author{Gabriel Ryan}
\authornotemark[1]
\affiliation{
  \institution{Columbia University}            
}
\email{gabe@cs.columbia.edu}          

\author{Todd Mytkowicz}
\affiliation{
  \institution{Microsoft Research}            
}
\email{toddm@microsoft.com}          

\author{Shuvendu K. Lahiri}
\affiliation{
  \institution{Microsoft Research}            
}
\email{shuvendu@microsoft.com}          

\begin{abstract}
Testing is widely recognized as an important stage of the software development lifecycle. Effective software testing can provide benefits such as bug finding, preventing regressions, and documentation. In terms of documentation, unit tests express a unit's intended functionality, as conceived by the developer. A test oracle, typically expressed as an condition, documents the intended behavior of a unit under a given test prefix. Synthesizing a functional test oracle is a challenging problem, as it must capture the intended functionality rather than the implemented functionality.

In this paper, we propose \toolName{} (a neural method for \underline{T}est \underline{O}racle \underline{G}ener\underline{A}tion), a unified transformer-based neural approach to infer both exceptional and assertion test oracles based on the context of the focal method. Our approach can handle units with ambiguous or missing documentation, and even units with a missing implementation. We evaluate our approach on both oracle inference accuracy and functional bug-finding. Our technique improves accuracy by 33\% over existing oracle inference approaches, achieving 96\% overall accuracy on a held out test dataset.
Furthermore, we show that when integrated with a automated test generation tool (EvoSuite), our approach finds \totalbugs{} real world bugs in large-scale Java programs, including \DISTINCTBUGS{} bugs that are not found by any other automated testing method in our evaluation.



\end{abstract}

\maketitle

\section{Introduction}
\label{sec:introduction}

Unit testing is a critical aspect of software development. Effective unit tests for a component (a method, class, or module) can provide documentation, find bugs, and prevent regressions. 
In terms of documentation, unit tests express the unit's \textit{intended} functionality, as conceived by the developer. 
Documenting the unit's functionality through a test conveys the unit's intended usage.
The test also serves as a mechanism for detecting functional bugs during development. 
When executed, a test checks for mismatches between intended and {\it implemented} functionality. 
Such a mismatch causes a test failure, indicating a bug in the implementation. 
Furthermore, unit tests can alert the developer when future code changes introduce bugs. 
Effective (unit) testing during development can prevent release of buggy software and reduce costs by billions of dollars \cite{national-testing-waste}.

A unit test is composed of two parts: a \emph{prefix}, which drives the unit under test to an interesting state, and an \emph{oracle}, which specifies a condition that the resultant state should satisfy. 
A sufficiently expressive test suite should document functionality under both normal invocations where the precondition is met, and {\it exceptional} behaviors where the precondition is violated.
Figure \ref{fig:pop_ex} shows two examples of unit tests for a \texttt{stack} class. The tests document a normal invocation (Figure~\ref{fig:pop_ex:cor}) and an exceptional invocation (Figure~\ref{fig:pop_ex:incor}). 
%
%
Figure~\ref{fig:pop_ex:cor} shows a normal invocation of the unit where the test prefix instantiates a \texttt{stack} and makes sequential calls to \texttt{push} and \texttt{pop}. 
The test oracle, highlighted in red, asserts that the stack's \texttt{isEmpty} method should return \texttt{true} at the resultant state.
If the unit contains a bug related to the tested behavior \eg{if pop always fails to remove an item from the stack}, this test can aid in detecting the bug. 
On the other hand, Figure~\ref{fig:pop_ex:incor} shows the unit's expected behavior when the precondition of \texttt{pop} is not satisfied. 
In this case, the intended behavior of calling \texttt{pop} on an empty \texttt{stack} is to raise an exception. 
As such, the test oracle is the expected exception. The try-catch structure ensures that the unit does indeed raise an exception.
If the unit contains a bug and does not raise an exception, the test will fail by executing \texttt{Assert.fail()}.

\begin{figure}[hb]
  \begin{subfigure}[b]{0.49\linewidth}
    \lstset{style=mystyle, xleftmargin=.0\textwidth, xrightmargin=0.0\textwidth, linewidth=0.95\linewidth}
\begin{lstlisting}[linewidth=4.1cm]
public void testPop() { 
  Stack<int> s = new Stack<int>(); 
  int a = 2;

  s.push(a);	
  s.pop();

  bool empty = s.isEmpty();
  !assertTrue(empty);!
}
\end{lstlisting}
      \caption{Normal invocation of \texttt{pop} \label{fig:pop_ex:cor}}
    \end{subfigure}
  \begin{subfigure}[b]{0.49\linewidth}
    \lstset{style=mystyle, xleftmargin=.04\textwidth}
\begin{lstlisting}[linewidth=4.2cm]
public void testPop() { 
 try {
  Stack<int> s = new Stack<int>(); 
  s.pop();
        
  Assert.fail(); //fail 
 } !catch (Exception e)! {
  //pass
 }
}
\end{lstlisting}
      \caption{Exceptional invocation of \texttt{pop} \label{fig:pop_ex:incor}}
    \end{subfigure}
    \caption{Unit tests of a \texttt{Stack} class. The test oracles are highlighted in red. A correct implementation of \texttt{Stack} will be empty after a sequential push and pop and must raise an exception if pop is called on an empty stack.}
    \label{fig:pop_ex}
\end{figure}

It is clear that testing has immense benefits. 
However, authoring high quality unit tests is time consuming. 
On average, developers spend 15\% of their time writing tests~\cite{testing-survey}. 
As such, extensive work has been devoted to automated unit test generation~\cite{EvoSuite, Randoop, Pynguin, AFL}. 
However, test generation tools have no definitive knowledge of the developer's intended program behavior. 
This creates a challenge for generating functional test oracles. 
Instead, these tools consider program crashes and undesirable exceptions (e.g. null dereference or out of bound array accesses) as the test oracles. 
These tests are capable of finding numerous {\it safety} bugs in the unit's implementation, but are not sufficient to find violations of intended functionality and thus do not replace the need for manual unit tests. 

Complimentary to automated test generation tools, extensive work has been devoted to test oracle creation from documentation and comments~\cite{alics, tcomment, JDoctor, C2S, toradocu}.
We refer to these techniques as {\it specification mining} methods for test oracle generation. 
These methods rely on a restricted structure of documentation and a set of handcrafted rules to infer exceptions and assertions for a unit.
However, given that users do not follow a prescribed format for writing documentation, or omit them altogether, these methods fail to extract interesting oracles on most real-world software components. 
In our evaluation, we show that these methods cannot infer bug-finding assertions for a benchmark of real world Java projects. 

Recently, neural generative models have shown promise in generating functional test oracles~\cite{AthenaTest, atlas, devdiv-assertions}. 
Neural methods are more flexible than specification mining approaches as they do not rely on fixed patterns.
This flexibility makes neural generative models robust to imprecise or even missing documentation. 
However, we find in our evaluation that these methods struggle to generate accurate oracles due to the large space of possible assertions. 

In summary, an effective test generation approach must infer both exception and assertion oracles that accurately reflect developer intent, and find bugs in real world programs. 
Additionally, such an approach must gracefully handle cases with ambiguous  or missing documentation, or even missing implementations. 

We propose a neural approach to infer both exceptional and assertion bug finding test oracles: \toolName{}. 
To address the limitations of existing neural generative methods, we propose a new approach that reformulates the oracle generation problem as a ranking over a small set of highly likely, possible oracles. 
We base our approach on the empirical observation that oracles in developer-written unit tests typically follow a small number of common patterns. 
We describe a taxonomy on these patterns and define a simple grammar that expresses this taxonomy. 
We use this grammar along with type-based constraints to restrict the space of candidate oracles and produce well-formed test oracles satisfying syntactic and type correctness. 
To perform ranking, we develop a two-step neural ranking procedure using pretrained transformers finetuned to score candidate oracles.

We evaluate our approach on both test oracle inference and bug-finding. 
Our technique improves accuracy by 33\% over existing oracle inference approaches, achieving 96\% accuracy on a held out test dataset that fits our grammar and constraints, and 69\% accuracy on an overall assertion benchmark, a relative improvement of 11\% over existing methods. Furthermore, we show that when integrated with a randomized test generation tool (EvoSuite), our approach finds \totalbugs{} real world bugs in Java benchmark, \texttt{Defects4J}~\cite{defects4j}. 
Our approach finds \DISTINCTBUGS{} bugs that are not found by any other automated testing method in our evaluation. We provide an open source implementation of \toolName{} at \repo{}.

\textit{Contributions.} In summary, this paper:
\begin{enumerate}
\item Introduces a transformer (neural network) based approach to generating both exceptional and assertion oracles without relying on the unit's implementation. 
\item Derives adapted datasets for exceptional and assertion oracle training that incorporate method signatures and docstrings. These datasets are included in our open source release.
\item Implements \toolName{}, an end-to-end test generation technique that integrates neural test oracle generation with the automated test generation tool, EvoSuite. 
\item Performs an extensive evaluation on test oracle inference. We demonstrate that our approach improves oracle inference accuracy by 33\% and finds \totalbugs{} real world bugs, including \DISTINCTBUGS{} bugs that are not found by any other method in our evaluation.
\end{enumerate}

    \section{Related Work}
\label{sec:related}

We broadly categorize related work on unit test generation into (i) automated test generation methods, (ii) specification mining methods, and (iii) neural methods. 

\begin{figure*}[t]
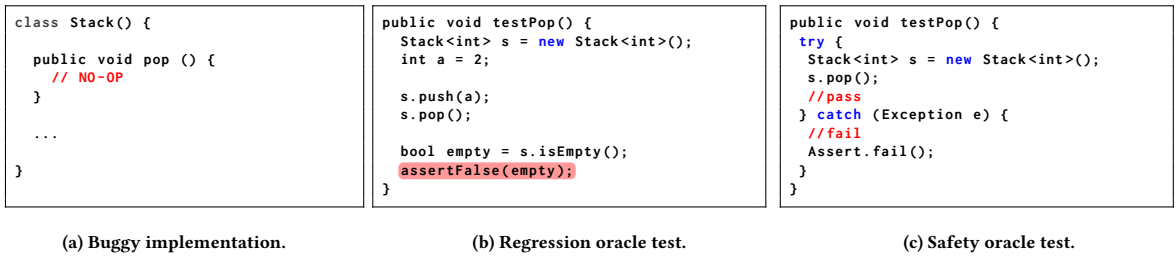

  \centering
  \begin{subfigure}[t]{0.3\textwidth}
    \lstset{style=mystyle, xleftmargin=.1\textwidth, xrightmargin=0.0\textwidth, linewidth=0.95\linewidth}
        \begin{lstlisting}
class Stack() {

  public void pop () {
    // NO-OP
  }
  
  ...
  
}
    \end{lstlisting}
    \caption{Buggy implementation.}
    \label{fig:ex_oracle:imp}
  \end{subfigure}
  \begin{subfigure}[t]{0.3\textwidth}
    \lstset{style=mystyle, xleftmargin=.0\textwidth, xrightmargin=0.0\textwidth, linewidth=0.95\linewidth}
\begin{lstlisting}[linewidth=5cm]
public void testPop() { 
  Stack<int> s = new Stack<int>(); 
  int a = 2;

  s.push(a);	
  s.pop();

  bool empty = s.isEmpty();
  !assertFalse(empty);!
}
\end{lstlisting}
    \caption{Regression oracle test. \label{fig:ex_oracle:regression}}
  \end{subfigure}
  \begin{subfigure}[t]{0.3\linewidth}
    \lstset{style=mystyle, xleftmargin=.0\textwidth, xrightmargin=0.0\textwidth, linewidth=0.95\linewidth}
\begin{lstlisting}[linewidth=5cm]
public void testPop() { 
 try {
  Stack<int> s = new Stack<int>(); 
  s.pop();
  //pass 
 } catch (Exception e) {
  //fail 
  Assert.fail();
 }
}
\end{lstlisting}
    \caption{Safety oracle test.\label{fig:ex_oracle:safety}}  
    \end{subfigure}
  \vspace{-5pt}
  \caption{Regression and safety oracles for a buggy pop method. The regression oracle (employed by EvoSuite) assumes that the current behavior is correct, resulting in an incorrect oracle asserting that the stack is non-empty. The safety oracle (employed by Randoop) assumes that any non-crashing behavior is correct. As such, it results in an incorrect oracle asserting that an exception should \textit{not} be raised when calling pop on an empty stack. Correct oracles for pop are shown in Figure~\ref{fig:pop_ex}.}
  \label{fig:ex_oracles}
  \vspace{-10pt}
\end{figure*}

\subsection{Automated Test Generation Tools}
\label{sec:related:random-test-gen}

Automated unit test generation techniques use a combination of black-box or white-box techniques to generate interesting \emph{test prefixes} for a unit. 
For example, tools such as Randoop~\cite{Randoop, randoop2} use random fuzzing of APIs of a unit to construct test prefixes that drives the unit to interesting states.
Fuzzers such as AFL~\cite{AFL} use fuzzing on the data inputs of a method to derive interesting values to drive a method.
Korat~\cite{Korat} performs test generation for data structure inputs based on lazy unfolding of the type structure. 
PeX~\cite{Pex} performs concolic execution~\cite{concolic,dart} to  enumerate paths in a program and synthesize inputs using a constraint solver to derive inputs. 

However, none of these tools explore the generation of \emph{test oracles} to find functional bugs in a unit.
They rely on program crashes (from implicit or explicit assertions present in the code), or use exception type heuristics to distinguish between desirable and undesirable behavior. For example, null dereferences or out of bounds exceptions may be considered as undesirable. 
\emph{Regression Oracles}, used by tools such as EvoSuite~\cite{EvoSuite, fraser2014large}, are intended to find \textit{future} bugs and assume the unit under test is correctly implemented. 
This assumption allows for generating assertions from observed execution behavior. 
However, expecting a correct implementation is not always a safe assumption. 
When the implementation is buggy, the regression oracles are incorrect with respect to the intended behavior.
That is, regression oracles are incapable of catching non-exceptional bugs, introducing false negatives. 

Consider the example in Figure~\ref{fig:ex_oracle:imp} that shows a buggy no-op implementation of stack pop. 
Figure~\ref{fig:ex_oracle:regression} shows a generated unit test with a regression oracle. 
The test creates a stack and makes sequential push and pop calls. 
Since the pop method has a buggy no-op implementation, the stack will have one element after executing pop. 
Thus, the regression oracle is an incorrect assertion: the stack should \textit{not} be empty. 

Similarly, qualifying any exceptional output as a bug (\emph{Safety Oracle}) can fail on correctly implemented methods, causing false positives, \eg{the intended behavior of calling \texttt{pop()} on an empty method is to throw an exception}. 
Figure \ref{fig:ex_oracle:safety} shows a generated unit test with a safety oracle. 
A method that relies on safety oracles will also generate a passing test on the buggy pop implementation. 
Since pop is implemented as a no-op, an exception will not be raised when calling pop on an empty stack. 
In this case, the test oracle is implicit and asserts that an exception will not be thrown. 

Therefore although automated test generation techniques find numerous non-functional bugs, and are useful for detecting regression bugs for future code changes, they are not a substitute for manually written unit tests documenting intended functionality. 



\subsection{Specification Mining Methods}
Specification mining works~\cite{alics, tcomment, JDoctor, C2S, toradocu} aim to generate test oracles that accurately reflect the intended behavior (as in Figure~\ref{fig:pop_ex}). Unlike randomized test generation methods, specification mining approaches do not have any knowledge of the unit's implementation and as such, do not require execution. Instead, they rely on docstring documentation. Specification mining methods typically define a set of natural language docstring patterns. These patterns cannot capture all docstrings as program comments can be written flexibly without any necessary syntax or structure.  

@Tcomment~\cite{tcomment} defines natural language patterns along with heuristics to infer nullness properties. However, it cannot generalize to other property or exception types. An example heuristic @Tcomment employs is: generate an ``expected NullPointerException'' oracle if the keyword \texttt{@param} has the words \texttt{null} and \texttt{not} within 3 words of each other.  ToraDocu~\cite{toradocu} uses a combination of pattern, lexical, and semantic similarity matching. Unlike @TComment, ToraDocu is not limited to nullness properties. However, ToraDocu can only generate oracles for exceptional behavior. JDoctor~\cite{JDoctor} is an extension of ToraDocu that can also generate assertion oracles. More recently, MeMo~\cite{BLASI2021111041} uses equivalence phrases in javadoc comments to infer metamorphic relations \eg{\texttt{sum(x,y) == sum(y,x)}}, which are also used as test oracles. These methods can precisely determine oracles when code comments fit their expected patterns, but do not generalize when comments fall outside these patterns.

Lastly, C2S~\cite{C2S} generates JML specifications from docstrings. C2S does not manually define patterns, but instead performs a search over JML tokens. However, C2S relies on a developer written test prefix to filter candidate assertions. C2S has performance improvements over JDoctor in terms of specification synthesis accuracy, but does not improve performance in bug finding.

On average, real-world Java projects lack precisely structured docstring documentation. In our evaluation, we show that specification mining methods struggle to infer bug-finding oracles for a benchmark of real world Java projects.

\parheader{Invariant Mining.} There is a long line of work in deriving program invariants for the observed execution behavior of the program. These include systems such as Daikon~\cite{ernst2007daikon} and DySy~\cite{csallner2008dysy}, which extends the derived program invariants with symbolic execution. Recently, EvoSpex~\cite{molina2021evospex, molina2020applying} combines observed executions with mutations to generate samples of both valid and likely invalid program states and applies a genetic algorithm to infer invariants for method postconditions. GAssert~\cite{gassert_fse2020} also utilizes an evolutionary approach to make inferred program invariants more accurate and compact.  These approaches can be used to generate specifications and associated test oracles from the inferred invariants, but because they are based on the execution/symbolic behavior of the current implementation they will generate regression oracles, and cannot detect if bugs are already present in the unit under test.


\subsection{Neural Methods} 
Recently, neural models have shown promise in generating test oracles and even entire unit tests. In contrast to specification mining methods, neural methods are not tied to hard coded patterns and can generalize to flexibly written docstrings. Furthermore, unlike randomized test generation tools, neural methods do not necessarily require knowledge or execution of the unit under test.

We refer the reader to CodeBERT~\cite{codebert} for a discussion on the transformer architectures as applied to code.  A transformer, like a recurrent neural network, maps a sequence of text into a high dimensional representation, which can then be decoded to solve downstream tasks.  While not originally designed for code, transformers have found many applications in software engineering~\cite{mergebert, gptc, pymt5, cubert}.

ATLAS is a neural-network-based approach to generate assertion oracles. Given a test prefix and the unit under test, ATLAS~\cite{atlas} generates assertions using a recurrent neural network. ATLAS relies on the unit's implementation and does not have any knowledge of the docstring documentation. ATLAS exclusively targets assertion oracle generation and does not attempt to infer any exceptional oracles.

Subsequent methods~\cite{devdiv-assertions, white2020reassert, mastropaolo2021studying} have improved upon ATLAS by using a transformer-based seq2seq architecture pretrained on natural language and code. A transformer seq2seq model outperforms ATLAS in terms of inference accuracy. However, in section~\ref{sec:eval}, we show that in combination with a test prefix generator, it struggles to find real world bugs in Java projects. 

Lastly, AthenaTest~\cite{AthenaTest} is a transformer model approach to generate entire unit tests including both prefixes and oracles. AthenaTest takes as input the unit's context \eg{surrounding class, method signatures, etc.}, and implementation. Like the previous neural methods, it does not have any knowledge of the docstring documentation and relies on the implementation for inferring intended behavior.

\section{Structure of an Oracle}
\label{sec:theory}

Our approach addresses the limitations of existing neural methods by employing a ranking architecture over a set of candidate test oracles, rather than a generative model.  In this section we develop a grammar for describing this set of test oracles. We first describe a taxonomy of commonly occurring oracle structures based on a qualitative investigation of a unit test dataset, and then use this taxonomy to inform the construction of our oracle grammar. 


We develop a taxonomy of common oracle structures based on unit tests from methods2test~\cite{methods2test}, a dataset of Java unit tests collected from GitHub. We describe methods2test in Section \ref{sec:4-exceptions} 


Unit test oracles typically test either exceptional behavior \ie{verifying an expected exception is raised} or return behavior (assertion oracles). Additionally, an implicit exception oracle is usually present in tests with assertion oracles. That is, a test with an assertion oracle is not expected to raise an exception. \\

\parheader{Taxonomy:} We develop the following taxonomy of oracle usage, drawn from our observations of almost 200K developer-written tests. To develop this taxonomy, we manually inspected 100 random samples and categorized the most frequently occurring types of oracles we observed. To ensure that our grammar generalized well and did not overfit to our 100 inspected samples, we evaluated the proportion of tests in the dataset that fit the grammar (Section \ref{sec:rq1}).

\begin{enumerate}
    \item \textbf{Expected Exception Oracles.} Expected exception oracles verify that executing the test prefix with some invalid usage raises an exception. They are most frequently expressed with the following structure:
    \vspace{-0.1cm}
    {\lstset{style=mystyle, frame=0, basicstyle=\linespread{0.75}\fontsize{7.8}{9.3}\ttfamily, keywordstyle=\color{blue}}
    \begin{lstlisting}
 try {
     Unit.methodcall(invalidInput);
     Assert.fail();
 } catch (Exception e) {
     verifyException(e, ExceptionType);
 }
    \end{lstlisting}
    }
    \item \textbf{Assertion Oracles.} Assertion oracles verify correct return behavior, although they will also fail if any exception is thrown. We observe several common assertion patterns:
    \begin{enumerate}
        \item \textbf{Boolean Assertions.} Boolean assertions are used to check some property of the unit under test is \texttt{true}/\texttt{false}. They are typically asserted directly on method return values:
{\lstset{style=mystyle, frame=0, basicstyle=\linespread{0.75}\fontsize{7.8}{9.3}\ttfamily, keywordstyle=\color{blue}}
    \begin{lstlisting}
 Unit.methodcall(input);
 assertTrue(Unit.getStatus());
    \end{lstlisting}
    }
        
        \item \textbf{Nullness Assertions.} Nullness assertions  usually check the return value of a method call that processes some input. 
{\lstset{style=mystyle, frame=0, basicstyle=\linespread{0.75}\fontsize{7.8}{9.3}\ttfamily, keywordstyle=\color{blue}}
    \begin{lstlisting}
 assertNotNull(Unit.processInput(input));
 assertNull(Unit.processInput(invalidInput));
    \end{lstlisting}
    }
        
        \item \textbf{Equality Assertions.} Developers typically write equality assertions to check the return value of a single method call. The return value is usually checked against a constant or literal representing the expected value. In many cases, especially when the unit under test incorporates some data structures, the expected value was previously passed as an argument to some method in the test prefix.
        {\lstset{style=mystyle, frame=0, basicstyle=\linespread{0.75}\fontsize{7.8}{9.3}\ttfamily, keywordstyle=\color{blue}}
    \begin{lstlisting}
 String msg = "foo";
 Unit.sendMessage(msg);
 assertEqual(Unit.getLastMessage(), msg);
    \end{lstlisting}
    }
    
    \end{enumerate}
\end{enumerate}
As we demonstrate in Section~\ref{sec:rq1} this taxonomy captures a majority of tests (82\% of a large dataset of developer written tests). This coverage could potentially be expanded by including other assertion types \eg{\texttt{AssertSame}}, however, in developing the oracle taxonomy, our goal is not to express the entire grammar of Java test oracles. Instead, we aim to identify a minimal syntactic subset which represents many semantically equivalent oracles. Such a grammar greatly restricts the output space for the oracle generator to consider.

\parheader{Uncommon oracles.} We note several other patterns that occur more rarely, including equality assertions on arrays or assertions on multiple method calls (as opposed to a method call and a constant).  We also note that there are some assertion patterns that we did not observe in \emph{any} unit test, although they are often used to express invariants within programs. These include assertions with logical connectives and assertions with inequality constraints.







\parheader{Test oracle grammar.} Based on the taxonomy of common oracle structures, we develop a restricted grammar that expresses commonly used test oracles. 

\vspace{10pt}
\hspace{-15pt}{\small\ttfamily
\begin{tabular}{@{}p{1.8cm}rp{0.15cm}l}
Test                &T &:=& \texttt{O(P)} \\
Prefix              &P &:=& \texttt{statement} | P; P \\
Oracle           &O(P) &:=& \texttt{E(P) | R(P)} \\
Except Oracle &E(P) &:=& \texttt{try\{P; fail();\} catch(Exception e)\{\}} \\
Return Oracle    &R(P) &:=& P; A \\
Assertion           &A &:=& assertEquals(const|var,expr) |\\
& & &assertTrue(expr) | assertFalse(expr) | \\
                         & & &assertNull(expr) | assertNotNull(expr) \\
\end{tabular}}\\ \\

Intuitively, \toolName{} is a code-generation model for tests that is explicitly designed to exploit the structure of a unit test. 
This grammar succinctly describes a set of test oracles that are possible candidates for generation. 
In particular, given a test prefix $P$, we can synthesize either an exceptional oracle $E(P)$ or an assertion oracle on the return value of a method $R(P)$.
Further the assertion oracle can be constructed using one of the five \texttt{assert*} constructs when instantiated with the return value and other constants and variables.

In the sections that follow, we demonstrate how to (i) prune this set, using type constraints, and (ii) rank the resulting possible test oracles using neural models.

\section{\toolName{}: Neural Test Oracle Generation} \label{sec:technique}

In this section we present our approach for inferring test oracles that reflect developer intent. Unlike previous works, \toolName{} is capable of inferring both exception and assertion oracles. Furthermore, \toolName{} can handle units with vaguely written or absent docstrings, or even absent implementation. Our approach infers test oracles from only a given test prefix and unit context. Unit context may refer to method signature(s), or a docstring (if present). Notably, the unit context need not include the unit's implementation. 

\subsection{Method Overview}

\begin{figure*}[htbp!]
    \includegraphics[width=\textwidth]{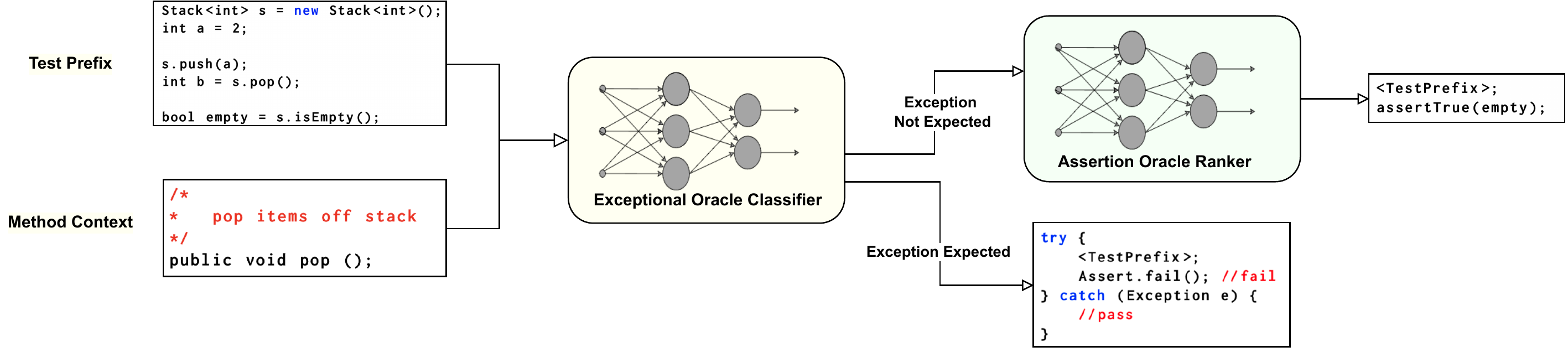}
    \caption{Overall \toolName{} framework. The system takes as input a test prefix and a unit context. The unit context contains method signature(s) and docstrings, but not the implementation. It outputs a unit test with an inferred test oracle. The system has two main components: the Exceptional Oracle Classifier and the Assertion Oracle Ranker.}
	\label{fig:our-framework}
\end{figure*}

\toolName{} depicted in Figure~\ref{fig:our-framework} contains two key components: the Exceptional Oracle Classifier and the Assertion Oracle Ranker.

The Exceptional Oracle Classifier, described further in Section~\ref{sec:4-exceptions}, is a pretrained transformer model fine-tuned on a binary decision task. The model decides if an exception should be thrown according to the developer intent conveyed through the unit context. If the classifier infers that the given test prefix should raise an exception, \toolName{} has found an exceptional oracle and can now generate a complete test. The resulting test has the \textit{Expected Exception Oracle} format shown in Section~\ref{sec:theory}. Otherwise, the classifier predicts that the input should not raise an exception and \toolName{} continues in the test generation process by invoking the Assertion Oracle Ranker.

The Assertion Oracle Ranker, described in Section~\ref{sec:4-assertions}, similarly uses a pretrained transformer model backbone. To address the limitations of existing neural assertion generation methods, our approach treats oracle inference as a ranking over a small set of possible common oracles. We base our approach on our observed taxonomy and defined grammar described in Section~\ref{sec:theory}. We use this grammar along with type-based constraints to restrict the space of candidate oracles and enforce syntactic and type correctness. The model is is fine-tuned on ranking the set of candidate assertions given the test prefix and unit context. Each assertion in the set is ranked, and the highest ranked candidate is selected as the assertion oracle. Lastly, \toolName{} generates a test with the given test prefix and the inferred assertion oracle.

\subsection{Exceptional Oracle Classifier}
\label{sec:4-exceptions}

As mentioned previously, the Exceptional Oracle Classifier is based on a pretrained BERT transformer model. In particular, we use the CodeBERT~\cite{codebert} model trained on both natural language and code masked language modelling. To train the Exceptional Oracle Classifier we fine-tune the pretrained model on the task of exceptional oracle inference. The fine-tuning is performed using a supervised dataset $D = {((p, c), l)_1, ... (p, c), l)_n})$ where $p$ is a test prefix, $c$ is a unit context, and $l$ is a binary label ($l \in {0,1}$). A label of 1 indicates that the sample should raise an exception while a label of 0 indicates that it should not raise an exception. 

\parheader{Methods2Test* dataset.} Our training dataset $D$ is variation of the Methods2Test dataset~\cite{AthenaTest}, we call Methods2Test*. As the name suggests, Methods2Test is a corpus of unit methods and corresponding developer written unit tests extracted from over 91K open source Java projects. Originally created to train AthenaTest, Methods2Test is structured for the translation task from methods to tests. We adapt Methods2Test to our setting of exception oracle inference. Our adapted dataset, Methods2Test*, has modifications in both the input methods and developer written tests. The input method's implementation is removed, and the method docstring (if present) is added. The tests are modified to remove any exception or assertion oracles. These stripped oracles are used to create binary labels for expected exceptions. Lastly, we normalize the test method name to prevent potential information leakage. For example, a test method named \texttt{testThrowsException} would leak label information to the model. To remedy this, we rename all tests to follow the format: \texttt{testN} where N is a positive integer. In summary, Methods2Test* is a supervised dataset for exception oracle inference. It excludes unit implementation and includes docstrings if present. Our resulting dataset contains a training set of more than 432,000 labeled samples.  
\subsection{Assertion Oracle Ranker} 
\label{sec:4-assertions}

The Assertion Oracle Ranker is also based on the pretrained CodeBERT~\cite{codebert} model. To train the Assertion Oracle Ranker we fine-tune the pretrained model on the task of assertion oracle inference. The fine-tuning is performed using a supervised dataset $D = {((p, c, a), l)_1, ... (p, c, a), l)_n}$ where $p$ is a test prefix, $c$ is a unit context, $a$ is a candidate assertion and $l$ is a binary label ($l \in {0,1}$). A label of 1 indicates that the given candidate assertion accurately reflects developer intent. For a given $p$ and $c$ only one $a$ can have a label of 1. The other assertions in the candidate set will have a negative label. 


\parheader{Atlas* dataset.} Our training dataset $D$ is a variant of the Atlas dataset~\cite{atlas}. Atlas is a corpus of test case prefixes, corresponding method units, and assertions. Atlas was collected from 9K open source Java projects on GitHub. We modify Atlas to create our variant dataset Atlas*. Similar to our construction of Methods2Test*, we remove the method implementation, normalize the test method name, and remove the assertion from the test case. Then, we generate a set of assertion candidates for each sample and construct our labels to indicate the correct assertion in the set. Our negative samples are also taken from the candidate set of assertions. In total the resulting Atlas* dataset contains over 170,000 labeled $(p, c, l)$ samples for supervised training.

\subsection{Candidate Assertion Set Generation}

To generate a candidate set of assertions, we use our grammar along with type-based constraints to restrict the space of candidate oracles and enforce syntactic and type correctness. Based on the return value of the unit under test, we iteratively construct a set of candidate assertions. Our assertion candidate generation algorithm is shown in Algorithm~\ref{alg:assertion_templates}. If the assertion that is being added requires an additional value (\texttt{assertEquals}), our approach draws likely candidates from Global and Local Dictionaries. 

\parheader{Global Constant Dictionary.} The Global Constant Dictionary contains the most frequently occurring constant values in the training data. Our global dictionary contains the top K values of each type. The use of a global dictionary is inspired by our observation that the vast majority of constants in test asserts are a few common values. For example, over 90\% of the integer constants in asserts in the ATLAS dataset are one of the top 10 most frequently occurring integer values.

\parheader{Local Dictionary.} In addition to the global constant dictionary, we also build a local dictionary based on values that appear in the test prefix. Note that these values are not necessarily constants. Variables that appear in the test prefix are also valid local dictionary entries. The use of a local dictionary is based on the observation that many assertions check against values that were previously passed as arguments to methods called in the test prefix.

At inference time, our method makes calls to the Assertion Oracle Ranker for each assertion in the set of candidates. The model outputs a predicted label along with a confidence score. We use this confidence score in post-processing to select the highest ranked assertion. The test prefix along with the selected assertion oracle is output as the generated test.

\begin{algorithm}
\caption{\label{alg:assertion_templates}Assertion Template Creation}
\begin{algorithmic}[1]
\Procedure{CreateCandidateTemplates}{GlobalDict, k, test}
\State $cs \gets \emptyset$ \Comment{Template Candidates}
\State retVal = extractRetVal(test)
\State t = type(retVal)
\State LocalDict = createLocalDict(test)
\If{retVal is an object} 
    \State $cs \gets cs \cup \texttt{assertNull(retVal)}$
    \State $cs \gets cs \cup \texttt{assertNotNull(retVal)}$
\ElsIf{retVal is a boolean}
        \State $cs \gets cs \cup \texttt{assertTrue(retVal)}$
        \State $cs \gets cs \cup \texttt{assertFalse(retVal)}$
\EndIf

\For {globalVal $\in$ GlobalDict.get(t)}
    \State $cs \gets cs \cup \texttt{assertEquals(globalVal, retVal)}$
\EndFor

\For {localVal $\in$ LocalDict.get(t)}
    \State $cs \gets cs \cup \texttt{assertEquals(localVal, retVal)}$
\EndFor

\State \Return $cs$
\EndProcedure
\\

\Procedure{CreateLocalDict}{test}
\State LocalDict = \{  \} 
\For {val in getValue(test)} \Comment{Loop over all values in prefix}
    \State LocalDict[type(val)] += \{val\}
\EndFor
\State \Return LocalDict
\EndProcedure 
\\

\Procedure{extractRetVal}{test}
\State assign = getLastLine(test) \Comment{last line will be an assignment}
\State retVal = getLHS(assign)
\State \Return retVal
\EndProcedure

\end{algorithmic}
\end{algorithm}

\subsection{End-to-End EvoSuite integration}
\label{sec:end-to-end-evo}
We have described a method, \toolName{}, to infer functional test oracles given a test prefix and unit context.
However, in order to catch bugs, a test prefix that exercises the buggy behavior is necessary. 
To obtain a high quality test prefix, we use the randomized test generation tool EvoSuite. As mentioned in Section~\ref{sec:related:random-test-gen} EvoSuite generates a set of tests guided by coverage. We extract test prefixes by stripping EvoSuite's oracles from each test. In cases where a test contains multiple assertions, we extract the test prefix for each assertion individually. For each of the generated test prefixes, we invoke \toolName{} to infer a test oracle. In combination with a large set of prefixes that attempt to cover the entirety of the unit, our approach is able to generate functional test oracles that find real world bugs.

When we obtain prefixes from EvoSuite, we assume that prefixes will be written in EvoSuite's standardized format. This allows us to identify the variables on which EvoSuite generated assertions in the \texttt{extractRetVal} method (Algorithm~\ref{alg:assertion_templates}). 

Lastly, we apply a confidence threshold to the assertion oracle ranker to suppress low confidence assertions. In these cases, only the exception oracle is applied to the test. Conceptually, this allows the model to avoid generating incorrect assertions in cases where the model believes all the candidate assertions are incorrect.

\section{Evaluation}
\label{sec:eval}

\parheader{Research Questions.} We consider the following research questions in our evaluation:
\begin{enumerate}
 \item[\textbf{RQ1}] Is \toolName{}'s grammar representative of most developer-written assertions?
 \item[\textbf{RQ2}] Can \toolName{} infer assertions and exceptional behavior with high accuracy?
 \item[\textbf{RQ3}] Can \toolName{} catch bugs with low false alarms? 
\end{enumerate}

\subsection{Evaluation Setup}

\parheader{Datasets.} Our evaluation uses the Atlas* and Methods2Test* datasets described in sections~\ref{sec:4-assertions} and ~\ref{sec:4-exceptions} respectively. For exceptional oracle inference, we evaluate on a Methods2Test* held-out test set of size 53,705. For assertion oracle inference, we evaluate on an Atlas* held-out test set of size 8,024.\\

\parheader{Bug Benchmark.} 
We evaluate real-world bug finding on the Defects4J~\cite{defects4j} benchmark. Defects4J is a benchmark of 835 bugs from 17 real world Java projects. Each sample in the benchmark includes both buggy and fixed code versions. Each fixed program version is based on a minimal patch to fix the bug, and passes all the project tests, while each buggy program version fails at least one test. Each bug is based on an error that was logged in the project's issue tracker, involves source code changes, and is reproducible \ie{with a deterministic test}. The benchmark also includes utilities for generating and evaluating test suites on the programs to determine if generated tests pass on the fixed versions and catch bugs. \\

\parheader{Test environment.} The evaluation was conducted on a Linux machine with Intel(R) Xeon(R) E5-2690 v3 CPU (2.60GHz) and 112GB main memory. As in the Defects4J environment, we use JDK 8. 

\subsection{RQ1: Oracle Grammar}
\label{sec:rq1}
We evaluate RQ1 on the original ATLAS dataset, which contains a total of 188,157 assertions mined from Java projects. 
To answer RQ1, we parse each assertion and check if it can be expressed in the grammar based on the assertion method name and structure of the AST. 
After excluding 695 samples that fail to parse, we find that 154,523 (82\%) can be expressed by our grammar. 

Of the 32,938 (18\%) of assertions that cannot be expressed in our grammar, the majority (23,913, 13\%) use  
assertion methods that we do not include \eg{\texttt{assertThat}, \texttt{assertSame}}. In many cases (74\% based on a manual inspection of 50 samples), the non-matching assertions appear to be symbolically equivalent to assertions expressible in our grammar (Figure~\ref{fig:mismatch-assert}). 

\begin{figure}[htbp!]
    \lstset{style=mystyle, xleftmargin=.05\textwidth}
\begin{lstlisting}[linewidth=8cm]
!assertThat(counter.get(),CoreMatchers.equalTo(2))!
vs.
@assertEquals(counter.get(),2)@
\end{lstlisting}
    \caption{The first assertion highlighted in red cannot be expressed in our grammar. However, the equivalent assertion highlighted in green, does fit our grammar.}
	\label{fig:mismatch-assert}
\end{figure}

Other assertions that did not match our grammar (5\%) include equality assertions on expressions rather than variables or literals. For example:
\begin{verbatim}
assertEquals(id1.hashCode(),id2.hashCode())
\end{verbatim}
Although we deliberately exclude generic assertions like these from our grammar, we note for a test executing in a deterministic environment, an equivalent property could be enforced through a syntactic rewrite.

\vspace{3pt}
\noindent \begin{longfbox}
{\textbf{Result 1:} 82\% of the developer-written assertions in the ATLAS dataset are in our grammar, and many other assertions are semantically equivalent to assertions expressed in our grammar.}
\end{longfbox}

\subsection{RQ2: Oracle Inference Accuracy}
To answer RQ2, Tables~\ref{tab:rq2-exceptions} and ~\ref{tab:rq2-assertions} reports accuracy results on a held-out test set. We include results for both exceptional and return test oracle inference. 

For exceptional oracle inference (Table~\ref{tab:rq2-exceptions}), our experimental setup involves the Methods2Test* dataset described in Section~\ref{sec:4-assertions}. There are no neural techniques for exceptional oracle inference that we are aware of. Instead, we include a random baseline (weighted coin) to illustrate the complexity of the problem space. The coin performs a random choice weighted on the distribution in our training set. In our training set, we observed that 80\% of samples are non-exceptional. As such, the coin predicts negative labels frequently (and usually correctly), but rarely predicts a positive. The coin performs similarly to our approach in terms of accuracy, but significantly worse in terms of F1 score, as it rarely predicts a positive label correctly. 

For assertion oracle inference (Table~\ref{tab:rq2-assertions}), our experimental setup involves the Atlas* dataset described in Section~\ref{sec:4-exceptions}. The accuracy metric is syntactic: a suggestion is considered correct if it is an exact lexical match. As a baseline, we compare to a sequence-to-sequence (seq2seq) return test oracle model~\cite{devdiv-assertions}.  The seq2seq model is a transformer pre-trained on natural language and code with a beam search decoder. In contrast to our approach which performs ranking over a set of template assertions, the seq2seq model generates a test oracle token by token. As such, the model suffers due to the large space of possible oracles. We report results on two held out test sets: an \textit{Overall} set and an \textit{In-Vocab} set. The in-vocab set is the subset of the overall set that can be expressed by our grammar and vocabulary based on the local and global dictionaries. 
Our model achieves 96\% accuracy on the in-vocab set compared to 63\% by the seq2seq model, and 69\% overall accuracy, an 11\% relative improvement over the seq2seq model.

\vspace{3pt}
\noindent \begin{longfbox}
{\textbf{Result 2:} Our assertion oracle inference model achieves over 69\% accuracy compared to 62\% accuracy from existing approaches. Our exceptional inference model achieves 86\% accuracy with an F1 score of .39 relative to a weighted coin baseline's .15 F1 score. }
\end{longfbox}
\vspace{3pt}

\begin{figure}[htbp!]
    \includegraphics[width=.5\textwidth]{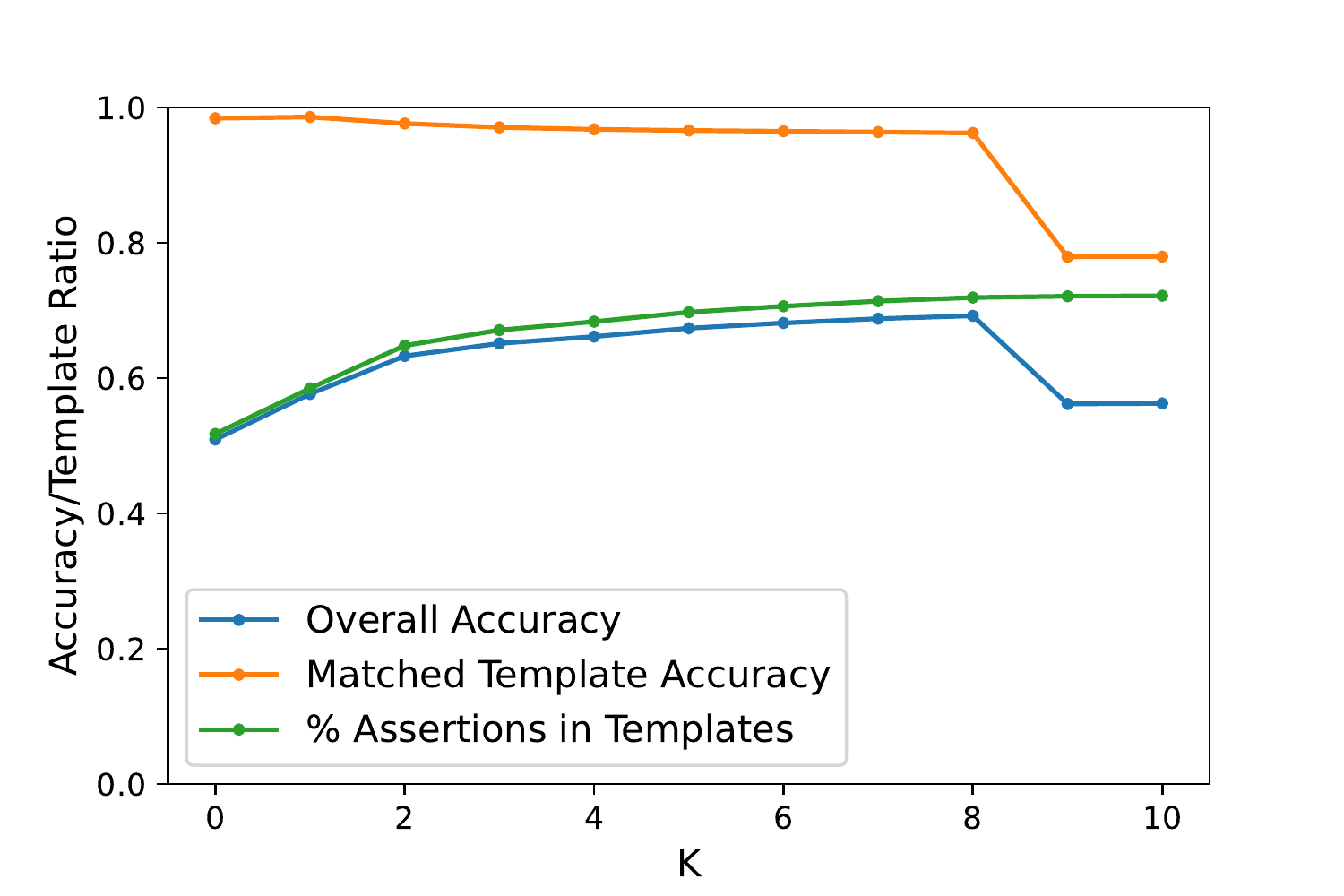}
    \caption{Evaluation of global dictionary size \texttt{K} on overall accuracy. Matched Template Accuracy indicates model accuracy when the candidate assertion set included the correct assertion. \% Assertions in Templates indicates the percentage of dataset assertions that appear in the candidate assertion set for a given \texttt{K}. }
    
	\label{fig:top-k}
\end{figure}

\parheader{Vocabulary size ablation.} We perform an study on \texttt{K}, the vocabulary size of our global dictionary, to examine the tradeoff between generating a larger number of assertion candidates and ranking the assertion candidates accurately. Figure \ref{fig:top-k} shows the overall model accuracy, percent of samples supported by the vocabulary, and accuracy on samples supported by the vocabulary evaluated on the ATLAS* test set. 

For \texttt{K}=0, the global dictionary is unused and only variables and constants in the local dictionary are considered the assertion generation. Using only the local dictionary can still generate correct assertion candidates for approximately 50\% of the samples in the test set. Increasing \texttt{K} causes the model accuracy to decline slightly, but causes overall accuracy to improve because more correct assertion candidates are generated using the global dictionary. Once the vocabulary becomes too large however, the model accuracy starts to drop off, and setting higher \texttt{K}s reduces overall accuracy.

In RQ2, we set \texttt{K}=8 based on tuning on the ATLAS* validation set. This setting achieves the best tradeoff between high model accuracy on the candidate set, and supporting a large set of likely assertions.


\begin{table}[t]
\resizebox{.45\textwidth}{!}{
{\small
	\begin{tabular}{ccccc}
	\toprule
		Approach & Accuracy & Precision & Recall & F1-Score \\
        \cmidrule(l){1-5}
		\toolName{} Model       & 86\%  & .55 & .30 & .39 \\
		Weighted Coin   & 76\%  & .15 & .13 & .15 \\
	 \bottomrule
	\end{tabular}
    }}
	\caption{RQ2: Evaluation of Exceptional Oracle Inference}
    \label{tab:rq2-exceptions}
\end{table}

\begin{table}[t]
\resizebox{.42\textwidth}{!}{
	\begin{tabular}{ccc}
	\toprule
		Approach                    & In-Vocab Accuracy       &  Overall Accuracy           \\
        \cmidrule(l){1-3}
        \toolName{} Model                   & 96\%                      &  69\%                       \\
		Seq2Seq    & 63\%                      &  62\%                       \\
	 \bottomrule
	\end{tabular}
    }
	\caption{RQ2: Evaluation of Assertion Oracle Inference}
    \label{tab:rq2-assertions}
\end{table}

\begin{table}[]
\centering
\resizebox{.37\textwidth}{!}{
\centering
    \begin{tabular}{ccc}
    \toprule
         Approach & Bugs Found (TPs) & FPR  \\
         \midrule
         EvoSuite + Ground Truth   & 120       & 0\%       \\
         EvoSuite + \toolName{} (Ours)   & \totalbugs{}        & 25\%  \\
         Randoop         & 20        & 87\%      \\
         EvoSuite + seq2seq & 6         & 46\%      \\
         AthenaTest     & 0         & $15\%^*$     \\
         EvoSuite + JDoctor        & 1         & 0.4\%      \\

         \bottomrule
    \end{tabular}
}
    \caption{RQ3: Overall Bug Finding. $^*$AthenaTest FPR based on 5 projects}
    \label{tab:rq3-overall} 
\end{table}

\begin{table}[]
\resizebox{.42\textwidth}{!}{
    \begin{tabular}{cccc}
    \toprule
                & Exception          & Exception       & Assertion   \\
         Approach       &  Raised          &  Not Raised       &  Failure  \\

         \midrule
         EvoSuite + Ground Truth   & 45                        & 27                          & 51                    \\
         EvoSuite + \toolName{} (Ours)   & 39                        & 5                           & 14                    \\
         Randoop         & 20                        & 0                           & 0                    \\
         EvoSuite + seq2seq  & 0                         & 0                           & 6                    \\
         AthenaTest     & 0                         & 0                           & 0                    \\
         EvoSuite + JDoctor        & 1                         & 0                           & 0                    \\

         \bottomrule
    \end{tabular}
}
    \caption{RQ3: Number of bugs found by oracle type. Note that some bugs can be detected by multiple oracle types. }
    \label{tab:rq3-breakdown}
\end{table}

\subsection{RQ3: Bug Detection}
To answer RQ3, we run our end-to-end test generation system, integrated with EvoSuite. As described in section~\ref{sec:end-to-end-evo}, the system uses EvoSuite to generate test prefixes guided by coverage. Our models are invoked to generate the test oracles.

\parheader{Baselines.} We consider the following baselines in this evaluation:
\begin{enumerate}
    \item \textbf{Randomized Test Generation.} To represent randomized test generation we run Randoop~\cite{Randoop}, which is a widely used and actively maintained test generation tool used for bug finding. We also run EvoSuite~\cite{EvoSuite} as a baseline, although EvoSuite's intended use case for regression testing limits its ability to find bugs present in the program. We run both Randoop and EvoSuite for 3 minutes per tested program, following the procedure used in ~\cite{shamshiri2015automatically}.
    \item \textbf{Neural Test/Oracle Generation.} To test neural methods, we compare with a seq2seq transformer finetuned to generate assertions~\cite{devdiv-assertions}. We also evaluate against a whole-test generation model, AthenaTest~\cite{AthenaTest}.
    \item \textbf{Specification Mining.} We use JDoctor's open source implementation to evaluate specification mining approaches. JDoctor supports exception oracle generation by parsing specific patterns in docstrings~\cite{JDoctor}. We integrate the generated oracles with the same EvoSuite-generated tests used by \toolName{} in this evaluation. Note that we do not evaluate on C2S~\cite{C2S} because the implementation is not publicly available.
\end{enumerate}

\parheader{Evaluation setting.} We evaluate RQ 3 on the Defects4J~\cite{defects4j} benchmark.  To evaluate the effectiveness of oracles in detecting bugs present in the program, the generated tests are run on a buggy version of the unit under test. We consider a bug is found if a generated test both fails on the buggy program and passes on the fixed program. Since each fixed program is distinguished from the buggy program by a minimal patch fixing the specific bug, a test must be failing due to the specific bug if it only fails on the buggy version.

For the oracle generation methods in the evaluation that require a test prefix (\toolName{}, seq2seq, JDoctor), we evaluate on a set of \emph{bug-reaching} EvoSuite test prefixes that exercise buggy behavior (and therefore can detect a bug given the right test oracle). We obtain this bug-reaching test prefix set by running EvoSuite with default settings \ie{coverage-guided} on the fixed program versions to generate regression tests, and then selecting tests that fail the buggy program version, indicating they exercise buggy behavior. We extract these tests' prefixes as an evaluation set. Methods evaluated on this set are denoted "EvoSuite + <method>" in Tables \ref{tab:rq3-overall} and \ref{tab:rq3-breakdown}.  

It is important to note that our evaluation setting is fundamentally different from the regression evaluation setting in which the Defects4J benchmark is most often used. In a regression evaluation, tests are generated on the \emph{fixed} program version and evaluated on the buggy version. Regression studies of randomized test generation tools report   finding larger numbers of bugs than in our setting as they use regression assumptions to generate higher quality oracles~\cite{shamshiri2015automatically}. In our setting where tests are generated on the buggy program version, regression test oracles will not find bugs as they assume the observed buggy behavior is correct.

In addition to evaluating the number of bugs found, we use per-test metrics as defined in~\cite{JDoctor}. These metrics include false positives to evaluate the performance of an oracle generation method from a usage perspective. A method that generates many erroneously failing tests will not usable in a realistic application setting where a developer must inspect each failure to determine if they represent real bugs or false alarms.

A failing test is considered a ``positive'' while a passing test is a ``negative''.  However, a ``positive'' does not necessarily indicate that the oracle caught the bug. A failing test can indicate one of two things:
\begin{enumerate}
\item \textbf{True Positive} - The test has a correct oracle and fails due to the buggy implementation.
\item \textbf{False Positive} - The test has an incorrect oracle and fails on the correct functionality of the unit in the fixed version.
\end{enumerate}
To distinguish between these cases, we run the same test on the unit's fixed version. If the test fails on the fixed version, we can safely assume the test has an incorrect oracle, and is a FP. 

Similarly, a passing test can indicate one of two things:
\begin{enumerate}
\item \textbf{True Negative} - The test has a correct oracle and is testing correct functionality.
\item \textbf{False Negative} - The test has an incorrect oracle and is testing buggy functionality.
\end{enumerate}
Again, to distinguish between these cases, we run the same test on the unit's fixed version. If the test fails on the fixed version, we can safely assume the test has an incorrect oracle, and is a FN.

\begin{figure}[htbp!]
    \includegraphics[width=.3\textwidth]{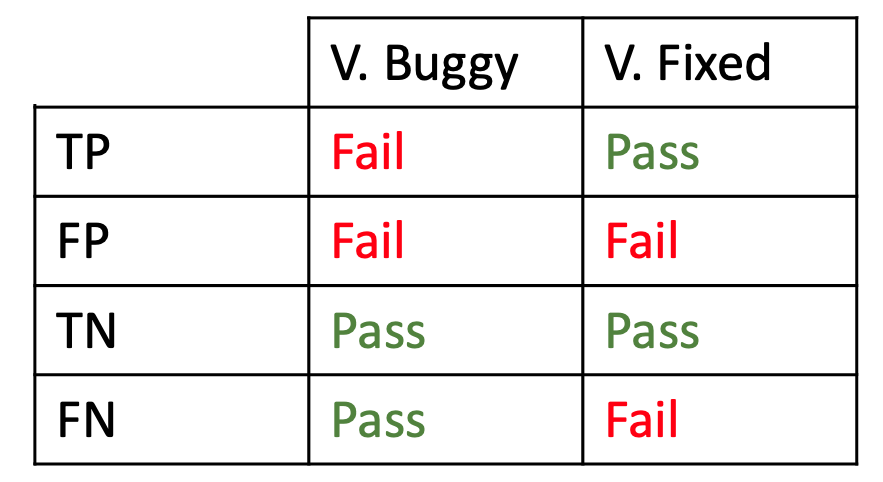}
    \caption{Bug Finding Metrics}
	\label{fig:metrics}
\end{figure}

We summarize the meaning of these metrics in Figure~\ref{fig:metrics}. In our evaluation, we summarize these metrics in the False Positive (FPR), which represents the rate of incorrectly failing tests on non-buggy code. A high FPR implies that a developer will need to validate many tests that have no utility and thus is a good metric for a bug-finding tool.

 \begin{figure*}[t]
  \centering
  \begin{subfigure}[b]{0.24\textwidth}
    \lstset{style=mystyle, xleftmargin=.0\textwidth, xrightmargin=0.0\textwidth, linewidth=0.95\linewidth}
    \begin{lstlisting}
class KeyedValues() {
  public void removeValue(int i){
    this.keys.remove(i);
    this.values.remove(i);
      
    // Bug, misses update
    if (i < this.keys.size()) {
      rebuildIndex();
    }
  }
  public int itemCount() {
    return this.index.size();
  }
}
    \end{lstlisting}
    \caption{Buggy implementation. \label{fig:keyed_values:imp}}
  \end{subfigure}
  \begin{subfigure}[b]{0.24\linewidth}
    \lstset{style=mystyle, xleftmargin=.0\textwidth, xrightmargin=0.0\textwidth, linewidth=0.95\linewidth}

    \begin{lstlisting}
public void testKeyedValues() {

  KeyedValues kv;
  kv = new KeyedValues(); 
  
  Short short = new Short(2); 
  kv.insertValue(0, short0, 2); 	
  kv.removeValue(0);
  
  // Asserts buggy itemCount 1
  // is correct and misses bug
  assertEquals(1,kv.itemCount());
}

    \end{lstlisting}
    \caption{Regression oracle test.\label{fig:keyed_values:reg}}  
    \end{subfigure}
  \begin{subfigure}[b]{0.24\linewidth}
    \lstset{style=mystyle, xleftmargin=.0\textwidth, xrightmargin=0.0\textwidth, linewidth=0.95\linewidth}

    \begin{lstlisting}
public void testKeyedValues() {
  try {
    KeyedValues kv;
    kv = new KeyedValues(); 
  
    Short short = new Short(2); 
    kv.insertValue(0,short0,2); 	
    kv.removeValue(0);
  
  // No exception raised
  } catch (Exception e) {
    fail(); // misses bug
  }
}
    \end{lstlisting}
    \caption{Safety oracle test.\label{fig:keyed_values:safety}}  
    \end{subfigure}
  \begin{subfigure}[b]{0.24\linewidth}
        \lstset{style=mystyle}

    \begin{lstlisting}
public void testKeyedValues() {

  KeyedValues kv;
  kv = new KeyedValues(); 
  
  Short short = new Short(2); 
  kv.insertValue(0, short0, 2); 	
  kv.removeValue(0);
  
  // Asserts itemCount should be 0
  // Test fails and identifies bug
  assertEquals(0, kv.itemCount());
}

    \end{lstlisting}
    \caption{\toolName{} generated oracle.\label{fig:keyed_values:func}}  
    \end{subfigure}
  \vspace{-5pt}
  \caption{Different types of test oracles for a bug in the \texttt{removeValue} method from the Java \texttt{Chart} project. The bug causes a data structure to return an incorrect item count when the most recently added item is removed. Although the test input exposes this behavior,  regression and safety oracles will generate a false negative by passing the buggy behavior, either by generating an incorrect assert statement or because the bug does not cause any exceptions to be thrown. Only the oracle generated by \toolName{} correctly asserts that itemcount should be 0 after an item is inserted and removed detects the bug. \toolName{} is the only system in our evaluation that correctly identifies this bug.}
  \label{fig:keyed_values}
  \vspace{-10pt}
\end{figure*}

\begin{figure}
  \begin{subfigure}[b]{0.49\linewidth}
    \lstset{style=mystyle, xleftmargin=.0\textwidth, xrightmargin=0.0\textwidth, linewidth=0.95\linewidth}

    \begin{lstlisting}
public void testStack() {
  try {
    NumberUtils.
      createNumber("0XT");
  
  // Safety Oracle
  } catch (Exception e) {
    fail();
  }
}


    \end{lstlisting}
    \caption{Safety oracle test.\label{fig:exception_oracles:safety}}  
    \end{subfigure}
  \begin{subfigure}[b]{0.49\linewidth}
        \lstset{style=mystyle}

    \begin{lstlisting}
public void testStack() {
  try {
    NumberUtils.
      createNumber("0XT");
    
  // Expected Exception
    fail("expecting exception");
  } catch (Exception e) {
    verifyException(e, 
      NumberFormatException);
  }
}
    \end{lstlisting}
    \caption{\toolName{} generated oracle.\label{fig:exception_oracles:functional}}  
    \end{subfigure}
  \vspace{-5pt}
  \caption{\label{fig:exception_oracles}Generated oracles testing a buggy \texttt{createNumber} method in the Java \texttt{Lang} project. The bug prevents a \texttt{NumberFormatException} from being raised on an invalid input. The oracle generated by \toolName{} correctly checks that an exception should be raised on the invalid input, and fails when no exception is raised due to the bug. A safety oracle cannot detect the absence of an exception. \toolName{} is the only system in our evaluation that detects this bug.}
  \vspace{-10pt}
\end{figure}

\parheader{Discussion of RQ3 Results.} Table~\ref{tab:rq3-overall} reports overall bug finding performance. EvoSuite + Ground Truth is a measure of EvoSuite's ability to generate bug-reaching tests. These tests were generated from the fixed program versions with regression oracles to obtain ground truth. We use this to distinguish between EvoSuite prefix generation performance from test oracle generation performance. EvoSuite + Ground Truth detects 120 bugs, indicating the best possible performance that any of the oracle generation methods can achieve on the EvoSuite test prefixes. 

\toolName{} finds \totalbugs{} total bugs, including \DISTINCTBUGS{} that are not found by any other method in our evaluation. 
The next best performing method, Randoop, finds 20 bugs but with a much higher false positive rate. Of the two tested neural methods, AthenaTest does not generate any bug-finding tests. The seq2seq model run on EvoSuite-generated test prefixes finds 6 bugs, but incurs a higher error rate. The specification mining tool, JDoctor, only finds one bug, but is the most precise oracle generation method in the evaluation. 
 
Table~\ref{tab:rq3-breakdown} reports a breakdown of bug finding performance on three different bug types: unexpected exception raised, expected exception not raised, and assertion failures. \toolName{}'s ability to infer exception oracles correctly is critical to its bug finding performance. Overall 44 of the bugs it finds are exceptional, and 5 involve expected exceptions not being raised. None of the other methods in the evaluation detect any expected exception not raised bugs. Of the other evaluated methods, AthenaTest and JDoctor are both capable of generating expected exception bugs in principle but in practice do not generate any in the evaluation. For raised (unexpected) exceptions, \toolName{} exception model correctly identifies 39/45 of them are unexpected exceptions. This demonstrates the value of using a neural model for exception oracle generation, which is more flexible than the fixed rules used by a tool like Randoop.

\toolName{} also identifies 14 assertion bugs. The only other method in the evaluation to generate assertion oracles that catch bugs is the seq2seq generative model, which catches 6 bugs. This shows that while \toolName{} ranking-based oracle generation procedure is effective for bug finding, its overall performance in bug finding comes from providing a unified method for oracle generation that can detect all three types of bugs. In contrast, none of the methods in the evaluation are successful in generating oracles for more than one type of bug, although JDoctor and AthenaTest can in theory generate oracles for all three classes of bugs.
 
The AthenaTest and seq2seq assertion generation models do not effectively find bugs. This evaluation illustrates the challenges in neural oracle generation. In practice we found that both AthenaTest's whole test generation and the seq2seq assertion model generated many tests and oracles that were not executable. The AthenaTest authors noted this issue in their evaluation, where they found that only 16\% of the generated test cases were executable without errors and actively tested the unit under test~\cite{AthenaTest}. The oracle generation model generated 34\% executable oracles, and of these we observed that a further 5\% were tautologies, resulting in an overall yield of 29\% potentially meaningful oracles. In contrast, the ranked oracle generation used by \toolName{} always generates oracles that are executable and exercise the unit under test. Note that due to the large volume of generated test candidates (30 per tested method) that must be individually compiled and run when following the procedure in ~\cite{AthenaTest}, we estimate the false positive rate of AthenaTest on five projects and otherwise only generate tests specifically on methods exercising buggy code.
 
The specification mining method, JDoctor also does not effectively find bugs, but it generates oracles precisely. JDoctor only produces an exceptional test oracle if there is a docstring comment indicating specific behavior. However, on the projects in the Defects4J benchmark, this approach only succeeds in generating test oracles to catch a single bug. We observed that in practice, many buggy methods either had vaguely worded docstrings or lacked docstrings entirely, and JDoctor created few test oracles as a result. JDoctor's inability to generate sufficient oracles to effectively find bugs illustrates why robustness to vague or missing docstrings is a important requirement for effective oracle generation. In many cases, the bugs detected by \toolName{} occurred on methods that lacked docstrings entirely, where any specification mining approach would not be able to identify them.

\parheader{EvoSuite vs. \toolName{} Performance:} Finding bugs requires both test prefixes that reach buggy behavior and oracles that correctly identify the bug. For the oracle generation methods in this evaluation, we distinguish the performance of the test prefix generator (EvoSuite) by evaluating the generated test prefixes with the ground truth oracles. Out of the 835 bugs in the Defects4J benchmark, the EvoSuite generated tests reach 120 bugs. That is, overall EvoSuite + \toolName{} misses 715 Defect4J bugs due to EvoSuite not generating reaching test prefixes, and 63 bugs due to \toolName{} not generating correct oracles. This result highlights that generating test prefixes to reach buggy code remains a challenging open problem, and improving the test prefix generator used with \toolName{} could have large impact on bug detection performance.

\parheader{\toolName{} Exception Oracle Error Analysis:} For a single focal method, EvoSuite often generates multiple test cases. For some focal methods (\textasciitilde 10\%), EvoSuite generates both exceptional and non-exceptional input states. However, \toolName{} rarely predicts (4\%) differing exception oracles for the same focal method, regardless of input state. This observation suggests that \toolName{} is conditioning primarily on the focal method signature rather than particular input states.

\parheader{\toolName{} Assertion Oracle Error Analysis:} We performed a manual analysis of ground truth oracles and found that of 229 total assertion oracles, 31 were predicted correctly. The remaining 198 predictions can be broken down as follows: 106 of the ground truth assertions could not be expressed with the given vocabulary, 13 could not be expressed with the grammar, and 9 were not predicted because \toolName{} incorrectly predicted an exceptional oracle. In the remaining 70 samples, the ground truth oracle could be expressed by the vocab and grammar, but the model made the wrong prediction, resulting in an in-vocab accuracy of 31\% on the bug-reaching EvoSuite tests. This is significantly lower than \toolName{}'s 96\% in-vocab accuracy on ATLAS*. The difference in performance suggests that the distribution of tests in ATLAS* is very different from EvoSuite's generated tests. A model trained specifically on EvoSuite generated (test, oracle) pairs instead of ATLAS* pairs may result in better performance.

\vspace{3pt}
\noindent \begin{longfbox}
{\textbf{Result 3:} Our approach finds \totalbugs{} bugs in real world Java projects, \DISTINCTBUGS{} of which are not found by any other method in the evaluation.}
\end{longfbox}

\subsubsection{Case Studies}

We consider two case studies of bugs that are detected by \toolName{} in our evaluation but not by other methods.

\parheader{Assertion bug case study.}  The first case study, shown in Figure \ref{fig:keyed_values} involves a bug in a key-value store used in the Chart Java project. The buggy method, shown in Figure \ref{fig:keyed_values:imp}, contains incorrect logic that prevents the data structure from updating its index when the most recently added item is removed. This causes the \texttt{itemCount()} method to return an incorrect count, because it bases the item count on the index. 

The EvoSuite-generated test for this method shown in Figure \ref{fig:keyed_values:reg} uses a regression oracle and generates an assertion based on the observed execution behavior. Because the method is buggy, this results in an incorrect assertion being generated, which not only fails to catch the bug but also could potentially make future detection of the bug more difficult. Figure \ref{fig:keyed_values:safety} shows a simplified version of an unexpected  exception oracle, which is the approach used by Randoop in the evaluation. 

In contrast to these two approaches, \toolName{} generates the correct oracle by performing a ranking over a small number of assertions on integers and the return value of \texttt{kv.itemCount()}. This identifies that after calling \texttt{removeValue(0)}, the most likely assertion is \texttt{assertEquals(0, kv.itemCount())}. 

\parheader{Expected exception case study.} Figure \ref{fig:exception_oracles} illustrates how \toolName{} is able to catch an expected exception bug detected in our evaluation. The bug in the \texttt{NumberUtils.createNumber} method of the Java \texttt{Lang} project prevents the method from correctly detecting invalid inputs and raising an exception.
The exception ranking model predicts that the \texttt{createNumber("0XT")} call should raise an exception based on the method signature and context, and \toolName{} generates an oracle based on this prediction to pass the test if an exception is raised on fail otherwise. In contrast, a safety oracle that checks for unexpected exceptions cannot detect this type of bug where a raised expected is desired behavior. 5 of the bugs found the \toolName{} in the evaluation are expected exception bugs, and no other tool in evaluation finds any expected exception bugs. 

\subsection{Threats to Validity}

We consider three potential sources of bias that could conceivably threaten the validity of our results: (i) test dataset bias, (ii) bug dataset bias, and (iii) bias from EvoSuite performance. Both of the unit test datasets ATLAS and Methods2Test sourced tests from publicly available Java projects, and filtered their results using heuristics such as GitHub star count and presence of matching focal methods to select tests for inclusion. Bias in these datasets towards specific applications or types of tests may effect the validity of RQ1 and RQ2. However, we note that these datasets are large (sourced from 91K open source Java projects in the case of Methods2Test), and therefore likely to be representative of common patterns in Java unit testing. 

Our RQ3 bug evaluation dataset, Defects4J, is much smaller with 835 samples from 17 projects due to difficulty in constructing minimal bug samples, so bias towards specific applications or bug types is possible. However, Defects4J only contains large, widely used projects and difficult real-world bugs, so evaluations on this benchmark are likely to be indicative of real world performance on large software projects. 

Finally, bias in EvoSuite's test prefix generation is also a potential threat to validity for RQ3. EvoSuite can only generate bug-reaching tests for a fraction of the Defects4J bugs (120 out of 835), and may be biased towards classes of bugs that are easier to reach with coverage guided exploration.

\section{Limitations}
\label{sec:limitations}

\textbf{Grammar and Vocabulary}: \toolName{} makes the tradeoff of supporting a restricted set of commonly used oracles, but predicting oracles in that set accurately. A limitation of this approach is that \toolName{} can only generate oracles that can be expressed by the grammar and exclusively contain values that appear in the vocabulary. We conducted a manual analysis of \toolName{} predictions in RQ3. When TOGA did not correctly predict a bug-finding assertion, in 54\% of the cases the assertion value did not appear in the vocabulary, and in 8.5\% of cases the assertion could not be expressed in the grammar. For example, \toolName{} could not predict the following ground truth assertion as the string literal is not contained in either the global or local dictionaries: \begin{verbatim} assertEquals("\"qDxD_5>q,)`dEgM", string0) \end{verbatim}
While our grammar limitation is strict, we found that approaches with unlimited vocabularies also did not correctly predict these oracles.

\noindent \textbf{Out of distribution training}: \toolName{} is also limited by its dependence on datasets of (primarily) developer-written unit tests for both training and vocabulary learning. However, the RQ3 test set is taken from EvoSuite, an out of distribution sample set. As a future direction, \toolName{} could be trained on an EvoSuite generated dataset for a model that more closely fits an end-to-end automated testing distribution. 

\noindent \textbf{Dependencies on EvoSuite}: \toolName{} assumes a particular structure of the test prefixes generated by EvoSuite to select the variable to assert on. However, as long as the assertion variable is specified to TOGA and defined somewhere in the test prefix, the test prefix could conceivably have any format. Therefore, integrating \toolName{} with another test generation method might require integrating a suitable mutation analysis tool such as PIT~\cite{PIT} to select variables on which to generate assertions.

\section{Conclusion}
\label{sec:conclusion}

This paper presents \toolName{}, a neural technique to infer both exception and assertion test oracles from a given test prefix and unit context. 
\toolName{} is a two step transformer based architecture that is capable of generating oracles for units without implementation or docstrings. It improves upon generative neural assertion oracle inference techniques by ranking a small set of likely candidate assertions. When integrated with a random test generation tool (EvoSuite) to obtain prefixes, \toolName{} finds \totalbugs{} real world bugs, out-performing existing test oracle inference techniques. Additionally, this paper presents two datasets for future work in neural exception and assertion test oracle inference.

\section*{Acknowledgements}

We would like to thank Michele Tufano and Alexey Svyatkovskiy for their help with the ATLAS and Methods2Test datasets and running AthenaTest, and helpful discussions and feedback. 



\bibliographystyle{ACM-Reference-Format}
\bibliography{refs}

\end{document}